\documentclass[conference]{IEEEtran}
\usepackage{cite}
\usepackage{multirow}

\usepackage{threeparttable}
\usepackage{booktabs}

\usepackage{graphicx}
\graphicspath{ {./images/} }

\ifCLASSINFOpdf
\else
\fi
\usepackage{amsmath}
\usepackage{amsfonts}
\usepackage{amssymb}
\usepackage[utf8]{inputenc}
\usepackage[caption=false,font=normalsize,labelfont=sf,textfont=sf]{subfig}

\usepackage{cite}
\usepackage{amsmath,amssymb,amsfonts}
\usepackage{algorithmic}
\usepackage{graphicx}
\usepackage{textcomp}
\usepackage{xcolor}
\usepackage{fancyhdr}
\usepackage{lipsum}
\def\BibTeX{{\rm B\kern-.05em{\sc i\kern-.025em b}\kern-.08em
    T\kern-.1667em\lower.7ex\hbox{E}\kern-.125emX}}

\makeatletter
\newcommand{\linebreakand}{%
  \end{@IEEEauthorhalign}
  \hfill\mbox{}\par
  \mbox{}\hfill\begin{@IEEEauthorhalign}
}
\makeatother

\makeatletter
 \let\old@ps@headings\ps@headings
 \let\old@ps@IEEEtitlepagestyle\ps@IEEEtitlepagestyle
 \def\confheader#1{%
 \def\ps@headings{%
 \old@ps@headings%
 \def\@oddhead{\strut\hfill#1\hfill\strut}%
 \def\@evenhead{\strut\hfill#1\hfill\strut}%
 }%
 \def\ps@IEEEtitlepagestyle{%
 \old@ps@IEEEtitlepagestyle%
 \def\@oddhead{\strut\hfill#1\hfill\strut}%
 \def\@evenhead{\strut\hfill#1\hfill\strut}%
 }%
 \ps@headings%
 }
 \makeatother

\begin{document}
%
\title{Automated skin lesion segmentation using multi-scale feature extraction scheme and dual-attention mechanism

}


\author{\IEEEauthorblockN{G Jignesh Chowdary}
\IEEEauthorblockA{\textit{Vellore Institute of Technology} 
\\
Chennai, India \\
jigneshchowdary@gmail.com}
\and
\IEEEauthorblockN{Ganesh V S N Durga Yathisha}
\IEEEauthorblockA{
\textit{V R Siddhartha Engineering College}\\
Vijayawada, India \\
gyathisha6807@gmail.com}
\and
\IEEEauthorblockN{Suganya G}
\IEEEauthorblockA{
\textit{Vellore Institute of Technology}\\
Chennai, India \\
suganya.g@vit.ac.in}
\linebreakand 
\IEEEauthorblockN{Premalatha M}
\IEEEauthorblockA{
\textit{Vellore Institute of Technology}\\
Chennai, India \\
premalatha.m@vit.ac.in}
}


%


\IEEEoverridecommandlockouts
\IEEEpubid{\makebox[\columnwidth]{ISBN: 978-1-6654-3811-7/\$31.00~\copyright2021 IEEE \hfill}
\hspace{\columnsep}\makebox[\columnwidth]{ }}
\maketitle
\IEEEpubidadjcol
\begin{abstract}
Segmenting skin lesions from dermoscopic images is essential for diagnosing skin cancer. But the automatic segmentation of these lesions is complicated due to the poor contrast between the background and the lesion, image artifacts, and unclear lesion boundaries. In this work, we present a deep learning model for the segmentation of skin lesions from dermoscopic images. To deal with the challenges of skin lesion characteristics, we designed a multi-scale feature extraction module for extracting the discriminative features. Further in this work, two attention mechanisms are developed to refine the post-upsampled features and the features extracted by the encoder. This model is evaluated using the ISIC2018 and ISBI2017 datasets. The proposed model outperformed all the existing works and the top-ranked models in two competitions.

\end{abstract}


%
\IEEEpeerreviewmaketitle

\section{Introduction}
Skin cancer is one of the leading causes of cancerous death around the world. It was estimated that nearly 13,29,779 new cases of skin cancer were diagnosed in 2018 \cite{I1}. Melanoma is the deadliest form of skin cancer, responsible for most skin cancer deaths. In 2018 \cite{I1}, the incidence of Melanoma was estimated to be 287700 with 60700 death cases. In spite of the high mortality rate of malignant Melanoma, it was observed that early-stage diagnosis could reduce the death rate and enhance the survival rate by five years on average for $95\%$ of the patients. But in later stages, the survival rate is as low as $15\%$, even with advanced medications and treatment procedures \cite{I2}.
Since Melanoma involves the pigmentation of lesions on the skin's surface, dermatologists can detect them through visual inspection. But this doesn't guarantee accurate and early-stage diagnosis in all cases. Besides this conventional method, Dermoscopy is the advanced non-invasive procedure for diagnosing Melanoma in the early stages. Dermoscopy eliminates the skin reflections and enhances the visualization ability of the deep skin, thereby enabling dermatologists/ oncologists to diagnose Melanoma which the human eye cannot see. Previous studies have shown that Dermoscopy enhanced the diagnostic accuracy of the conventional procedure \cite{x}. But the manual examination of these dermoscopic images is a non-reproducible and time-consuming process due to the complexity of the lesions and an immense number of images \cite{y}. Manual examinations often contain considerable human error, resulting in a faulty diagnosis \cite{z, q, w}. For several cases, it is even difficult for experienced medical professionals to diagnose Melanoma from dermoscopic images due to the varying characteristics of the tumor. There was a great interest in developing Computer-Aided-Diagnostic (CAD) systems for assisting medical professionals in clinical evaluation \cite{I3}. One of the important components for developing such CAD systems is the automatic segmentation of lesions from dermoscopic images so that these regions can be used for further analysis \cite{I4}. Designing such automated segmentation methods is challenging due to the variations in shape and size, irregular lesion boundaries, and minimum contrast difference between the lesion and the skin. Researchers have worked a lot to resolve these issues.
Early segmentation approaches employed edge detection, region growing, and optimum thresholding methods. Saez et-al. \cite{I5} used an edge-based-level-set algorithm for the segmentation of the lesion. Grana et al. \cite{I6} utilised the Catmull-Rom spline approach to predict the lesion boundary after calculating the lesion slope regularity and slope numerically. Lesions with less contrast difference between the skin can be better diagnosed by combining several threshold-based methods than single-threshold approaches \cite{I7}. For finding the optimal colour channel for lesion segmentation, some research works \cite{I8, I9} used cluster-based histogram threshold and colour space analysis. These traditional approaches require several hyperparameters to be fine-tuned for achieving high segmentation performance.

\par With the rapid developments of the convolutional neural networks (CNN) for medical image segmentation \cite{TMS}, several studies employed them for skin lesion segmentation. Ghafoorian et-al. \cite{I10} proposed a multi-branch DCNN for extracting multi-scale context features, but this network is too shallow to extract high discriminative features. With the development of batch-normalization \cite{I12}, and residual \cite{I11}, the problem of network degeneration and the vanishing gradient is solved, resulting in the networks going deeper. Yu et al. \cite{I13} reported that the deep architectures could extract high discriminative features for skin lesion segmentation, but these networks neglected global features as they are focussed on local context, restraining them from achieving more accurate results with deep architectures. 

Recently, attention mechanisms have become popular in deep learning to extract global features for accurate segmentation. In the study \cite{I14}, the attention mechanism coupled with the popular UNet architecture is used to select the discriminative features for different organs with varying characteristics like size and shape by weighting the different channels. But these single attention approaches fail in cases of lesions with complex characteristics. 

In this work, we proposed an UNet based segmentation framework that employs a novel multi-scale feature extraction module for extracting highly discriminative deep features for skin lesion segmentation. In addition to the multi-scale feature extraction module, we also employed a dual attention mechanism for refining the post-upsampled features and the features extracted by the decoder. This work is experimented using the ISBI2017 and ISIC2018 datasets. The experimental results reported that the proposed model achieved better segmentation performance than the existing state-of-the-art works.

\section{Related works}
Over the last decade, several researchers have proposed computerized approaches for the segmentation of skin lesions. These methods can be categorized into active contour models, region growing and splitting, thresholding, clustering, and supervised learning like Celebi et al. \cite{M1} employed the ensemble of four thresholding-based methods for the estimation of the lesion boundary. Peruch et al. \cite{M2} proposed a novel two-stage approach named Mimiking-Expert-Dermatologists-Segmentation (MEDS) to segment the skin lesions. In the first stage, the Principle Component Analysis (PCA) is employed to project the skin lesion image onto the first principle component of the color-histogram, and in the second stage, the thresholding method is applied to mimic the cognitive procedure of the human dermatologist by clustering the pixels based on color into non-lesion and lesion regions. These thresholding-based approaches mainly depend on the histogram of the image color; thus, these methods suffer when the images contain a significant amount of bubbles, hair, or other unwanted structures. To overcome the effect of such distractions, Zhou et al. \cite{M3} employed the classic GVF Snake algorithm and included a mass density function into the optimization objective functional,  which may be solved via mean shift estimation. However, this optimization procedure requires a huge amount of computation for achieving convergence. Sadri et al. \cite{M4} employed a fixed-grid wavelet network for segmentation. In this work orthogonal least squares method was employed to supervisely optimize the network topology and for the calculation of the network weights. Xie et al. \cite{M5} coupled the generic algorithm with the self-generating neural network for the accurate segmentation of skin lesions. All these supervised and the combination of unsupervised and supervised approaches require particularised domain knowledge as they rely on hand-crafted features. 
\par With the rapid developments in deep learning, deep convolutional neural networks (DCNN) have become more popular in solving computer vision tasks \cite{lw1, lk1}. These DCNN's have the ability to extract highly discriminative hierarchical features from raw images, thereby eliminating the need for hand-crafted features. Besides the success of these models in natural image classification tasks \cite{M6}, they have also shown promising results in medical imaging like the diagnosis of Mitosis from histopathology images \cite{M7}, skin cancer classification \cite{M8}, and image registration \cite{M9}. In addition to medical image classification, the DCNN's have shown excellent results in the segmentation of tumors from MRI images \cite{M10}, left ventricle from cardiac MRI images \cite{M11}, amongst others. With this motivation, these methods are also used for skin lesion segmentation from non-dermoscopic images \cite{M12}. But all these segmentation approaches employed patch-passed classification, in which the input image is divided into several patches, and then each patch is classified/predicted as outside the target or within the target. This technique only integrates limited contextual information included in the patch because each patch only depicts a local region of the image. The contextual information can be enhanced by enlarging the parches, but large patches result in the loss of fine details, which play a major role in final segmentation. The sliding window approach can be employed to integrate both global and local contextual information in segmentation. However, this approach is highly computationally expensive and is not efficient due to overlapping patches.

\par Due to the limitations in patch-based and sliding window approaches, few researchers employed DCNN's with global dermoscopic images for skin lesion segmentation. Yu et al. \cite{I13} developed a neural network with an embedded residual module to enhance the feature extraction capability of the model for efficient skin lesion segmentation. Yuan et al. \cite{M13} proposed a new loss function based on Jaccard Similarity for optimizing the segmentation task. Sarkar et al. \cite{M14} used the combination of four pre-trained networks in the encoder for extracting discriminative features and proposed a novel loss function based on softmax. These pre-trained networks are extended by pyramid-pooling modules. For a stronger feature representation, Alom et al. \cite{M15} presented a novel recursive-residual layer to extract features based on cyclic convolutions. Even though these methods have shown satisfactory performance by enhancing the capability to extract local features, they fail to extract required global features for significantly higher performance. Recently by combining multiple backbones, Jahanifar et al. \cite{M16} proposed an ensemble approach for the segmentation of skin lesions. But the ensemble of multiple models will increase the number of parameters, and such models require more run-time for network convergence. Hence such an approach is complex and difficult to be employed in a clinical scenario.
Traditional DCNN's ignore long-range dependencies as they don't have any smart mechanism to guide feature selection. Wang et al. \cite{M17} used a non-local block based on an attention mechanism to represent long-range pixel-wise relationships in order to get long-range dependencies. Hu et al. \cite{M18} employed a $Squeeze_and_Excitation$ (SE) block for modeling global context information, which re-calibrated channel dependencies by scaling distinct channels. However, these single attention mechanisms are incapable of dealing with the challenges in skin lesion segmentation due to the complex and variant characteristics of lesions. In this work, a new feature extraction module and dual attention mechanism are employed for faster and efficient segmentation of skin lesions.

\section{Materials and Methods}
\subsection{Residual Multi-Scale Module (RMSM)}
\begin{figure}
    \centering
    \includegraphics[width=\columnwidth]{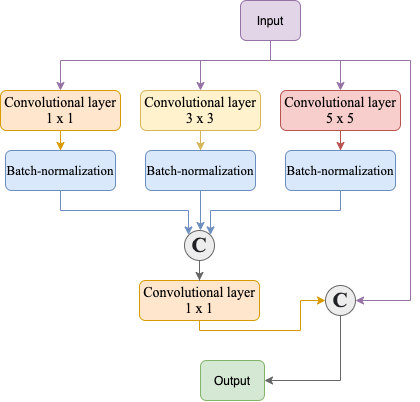}
    \caption{Multi-scale feature extraction module with residual connection (RMSM). \label{msm}}
    
\end{figure}

In this module, multiple convolutional layers with varying kernel sizes are used to extract multi-scale information for each pixel. This multi-scale method enhances the segmentation performance, as the large scale provides more spatial information, whereas smaller scales give more comprehensive information about each pixel's immediate neighbors \cite{W2}. Moreover, the residual connection makes learning easier for the network. The structure of the proposed RMSM is shown in Figure \ref{msm}. In this module, the batch-normalization layer is used after each convolutional layer except for the bottleneck layers to avoid the problem of vanishing gradients while retaining convolutional layers. The ourtput of the RMSM can be computed using Equation \ref{euotp}.

\begin{equation}
    \begin{aligned}
        Out_{RMSM} = I_n \oplus Cov_{1 \times 1}(L_{BN}(Cov_{1\times1}(I_{n})) 
        \\\oplus L_{BN}(Cov_{3\times 3}(I_n))  \oplus L_{BN}(Cov_{r\times 5)}(I_n))) 
        \label{euotp}
    \end{aligned}
\end{equation}

Where in Equation\ref{euotp}, $Out_{RMSM}$ represents the output feature map of the RMSM, $I_n$ represent the input feature map, $L_{BN}$ represent the batch normalization layer, $oplus$ represents the concatenation, and $Cov_{1\times1}$, $Cov_{3\times3}$, $Cov_{5\times5}$ represent convolutional layers with respective kernel sizes.

\subsection{Attention mechanisms}
\subsubsection{Decoder feature refinement attention module (DF-RAM)}
Feature concatenation between the encoder and decoder at each stage of the network is the topological structure of the UNet. The feature maps from the encoder are combined with upsampled feature maps of the decoder for the better localization of the intended segmentation targets \cite{W1}. But, not every visual representation of the encoded features helps for accurate segmentation. Besides this, the semantic gap between the decoder and encoder may hinder the segmentation performance. So an attention module (DF-RAM) is employed for the refinement of encoder features before concatenation. The DR-RAM refines features using both channel and spatial attention; the structure of DF-RAM is shown in Figure \ref{de}. 
This module requires a $Decoder\_Feature (DF)$ and $Skip\_Feature (SF)$, where $Decoder\_Feature$ refers to the feature map from the last decoder block (or feature map from the previous encoder block) and $Skip\_Feature$ refers to the feature map from the encoder block at the same level fed through the skip connection. The $DF$ is denoted by $D \epsilon \mathbb{R}^{N_{d}\times L_{d} \times K_{d}}$ and the $SF$ is represented as $S \epsilon \mathbb{R}^{N_{s}\times L_{s} \times K_{s}}$.

In comparison to $SF$, $DF$ contains semantic information contained in the channel dimension. So in this work, a Max-Pooling operation followed by a Multi-layer perceptron is used for creating the channel-attention map $CH_{A}(S)\epsilon \mathbb{R}^{N_{s}\times 1 \times 1}$. The size of the output of the MLP is smaller than the input; this removes the irrelevant information from the channel dimension. The channel attention is presented in Equation\ref{ca}. 

\begin{equation}
    CH_{A}(S) = \sigma(MLP(P_{GAP}(D))
    \label{ca}
\end{equation}

In Equation\ref{ca}, $\sigma$ represents the sigmoid activation, and $P_{GAP}$ represents the Global Average Pooling.

In spatial attention, both $DF$ and $SF$ are used. Firstly a convolutional layer with $1\times 1$ filter is employed to reduce the channel dimensions of $SF$ and $DF$. Then the squeezed $DF$ is upsampled to complement the size of $SF$ for channel-wise concatenation. These concatenated features are passed through a set of two convolutional layers with variant kernel sizes for producing spatial attention $SP_{A}(S)\epsilon \mathbb{R}^{1 \times L_{s} \times K_{s}}$. The spatial attention is presented in Equation5.

\begin{equation}
\begin{aligned}
    SP_{A}(S) = \sigma(Con_{3\times3}([S_{r}, D_{r}])+Con_{5\times5}([S_{r}, D_{r}])+\\Con_{7\times7}([S_{r}, D_{r}])+Con_{9\times9}([S_{r}, D_{r}]))
    \label{rty}
\end{aligned}    
\end{equation}

Here $D_{r} = Up\_sample(Conv_{1\times1}^{r}(D))$, $S_{r} = Conv_{1\times 1}^{r}(S)$

In Equation\ref{rty}, $r$ is the reduce ratio and is set to 16 in this work, $Conv_{3\times3}$, $Conv_{5\times5}$, $Conv_{7\times7}$, $Conv_{9\times9}$ represent the convolution operations with respective $kernel\_sizes$, and the $Conv_{1\times1}^{r}$ is employed to squeeze the channel dimension.
Finally, element-wise multiplication is employed for combining spatial and channel attention for generating fused attention $DF(S)$, shown in Equation \ref{de}.

\begin{equation}
    DF(S)=S\otimes CH_{A}(S) \otimes SP_{A}(S) 
    \label{de}
\end{equation}

In Equation \ref{de}, $\otimes$ represents the $element\;wise\;multiplication$.

\begin{figure}
    \centering
    \includegraphics[width=\linewidth]{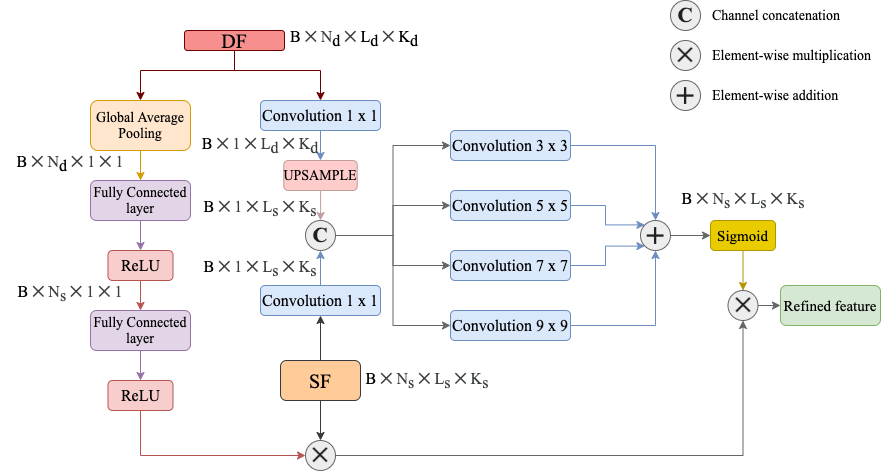}
    \caption{Proposed Decoder Feature Refinement Attention Module (DF-RAM). It uses $SF$ and $EF$ for refining the encoder features. Here $B$ represents the batch size, $N$ represents the number of channels, $L$, and $K$ represent the height and width of the feature maps.\label{de}}
    
\end{figure}

\subsubsection{Encoder feature refinement attention module (EF-RAM)}
\begin{figure}
    \centering
    \includegraphics[width=\linewidth]{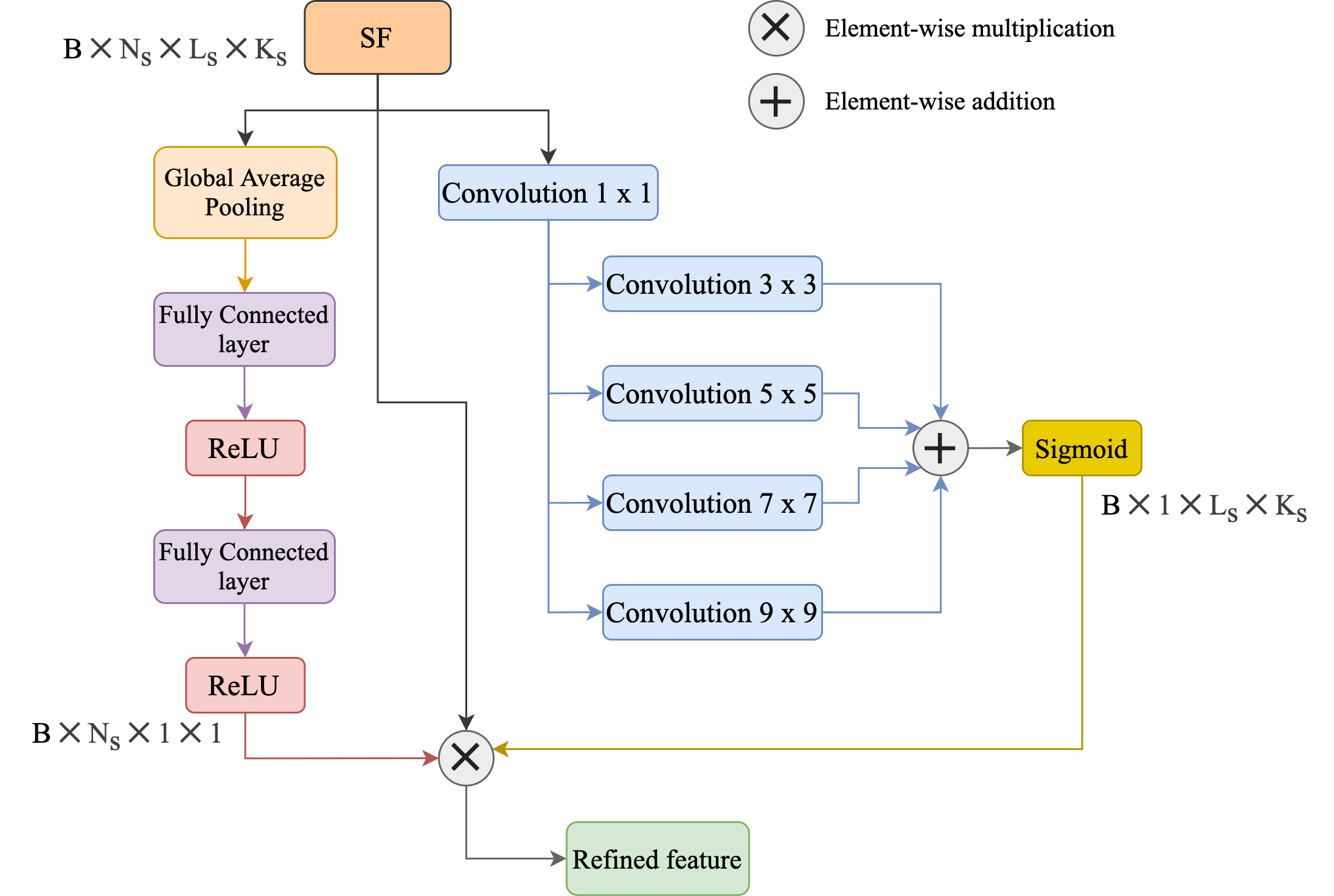}
    \caption{Proposed Encoder Feature Refinement Module (EF-RAM). It uses $SF$ to refine the post-upsampled feature maps. Here $B$ denotes the batch size, $N$ denotes the number of channels, $L$, and $K$ denotes the height and width of the feature maps. }
    \label{teop}
\end{figure}

For semantic segmentation, the visual representation extracted by the encoder needs to be upsampled for making dense predictions. Interpolation and transposed convolutions are the two approaches for image upsampling, although each has its own set of limitations. Compared to interpolation, transposed convolutions enhance the model capacity as they are trainable and offer non-linearity to the segmentation models. However, improper fine-tuning of hyper-parameters increases the grid effects, and this becomes more complex when more than one transposed convolutions are stacked. Thus, in this work, bi-linear interpolation with the following convolution is employed. As interpolation is non-trainable, noise and other irrelevant information are possible during the upsampling process. To overcome this problem, an attention mechanism is introduced in this work. The attention mechanism refines the feature maps that have been upsampled in both spatial and channel dimensions, as shown in Figure \ref{teop}. The spatial attention is denoted by $SP_{A}(S)\epsilon \mathbb{R}^{N \times 1 \times 1}$ and the channel attention is represented by $CH_{A}(S)\epsilon \mathbb{R}^{N \times 1 \times 1}$.The channel and spatial attentions can be computed using equations \ref{chop} and \ref{poelr}. 

\begin{equation}
    CH_{A} = \sigma(MLP(P_{GAP}(S))))
    \label{chop}
\end{equation}

\begin{equation}
    \begin{aligned}
        SP_{A}(S) = \sigma (Con_{3 \times 3}(Con_{1 \times 1}^{r}(S))+Con_{5 \times 5}(Con_{1 \times 1}^{r}(S))+\\Con_{7 \times 7}(Con_{1 \times 1}^{r}(S))+Con_{9 \times 7}(Con_{1 \times 1}^{r}(S)))
    \end{aligned}
    \label{poelr}
\end{equation}

In Equation \ref{chop} and \ref{poelr}, $r$ is the reduce ratio and is set to 16, $P_{GAP}$ represents the Global Average Pooling, $\sigma$ denotes the sigmoid activation, $Con_{3 \times 3}$, $Con_{5 \times 5}$,  $Con_{7 \times 7}$, and  $Con_{9 \times 9}$ represents the convolutional operations with respective kernel sizes. And $Con_{1 \times 1}^{r}(S)$ is used for squeezing the channel dimension. The fused attention $EF(s)$ is computed 

The fused attention $EF(S)$ is computed by combining $SP_{A}(S)$ and $CH_{A}(s)$ through element wise  multiplication $\otimes$, shown in Equation \ref{lfppg}

\begin{equation}
    EF(S) = SP_{A}(S)\otimes CH_{A}(S)\otimes S
    \label{lfppg}
\end{equation}

\subsection{Residual Multi-Scale dual attention Unet}
In this work, a 4 level architecture of Residual Multi-scale Dual Attention UNet is proposed to segment skin lesions from dermoscopic images. This network consists of three paths: encoder, decoder, and bridge. The encoder path extracts the compact representations from the input images, the decoder path recovers the extracted representations to the pixel-wise segmentation, and the bridge connects the encoder and decoder paths. The encoder path consists of three RMSM modules. In each RMSM module, a stride of 2 is applied to the last convolutional layer for downsampling the feature map. The decoder path consists of three RMSM modules, and before each module, there is an upsampling of the feature map from the previous module and concatenation from the RMSM module in the encoder path at the corresponding level. The bridge consists of one RMSM module. In this network, DF-RAM and EF-RAM modules are inserted for enhancing the segmentation performance. The DF-RAM module is used to refine the feature extracted by the encoder and minimize the segmentation gap and is inserted before the concatenation. And the EF-RAM module is inserted in the decoder path after each RMSM module for refining the upsampled feature maps in both spatial and channel dimensions. At the end of the decoder path, a $1 \times 1$ convolutional layer followed by a sigmoid activation layer is used for producing the segmented output.

\begin{figure*}
    \centering
    \includegraphics[width=\textwidth]{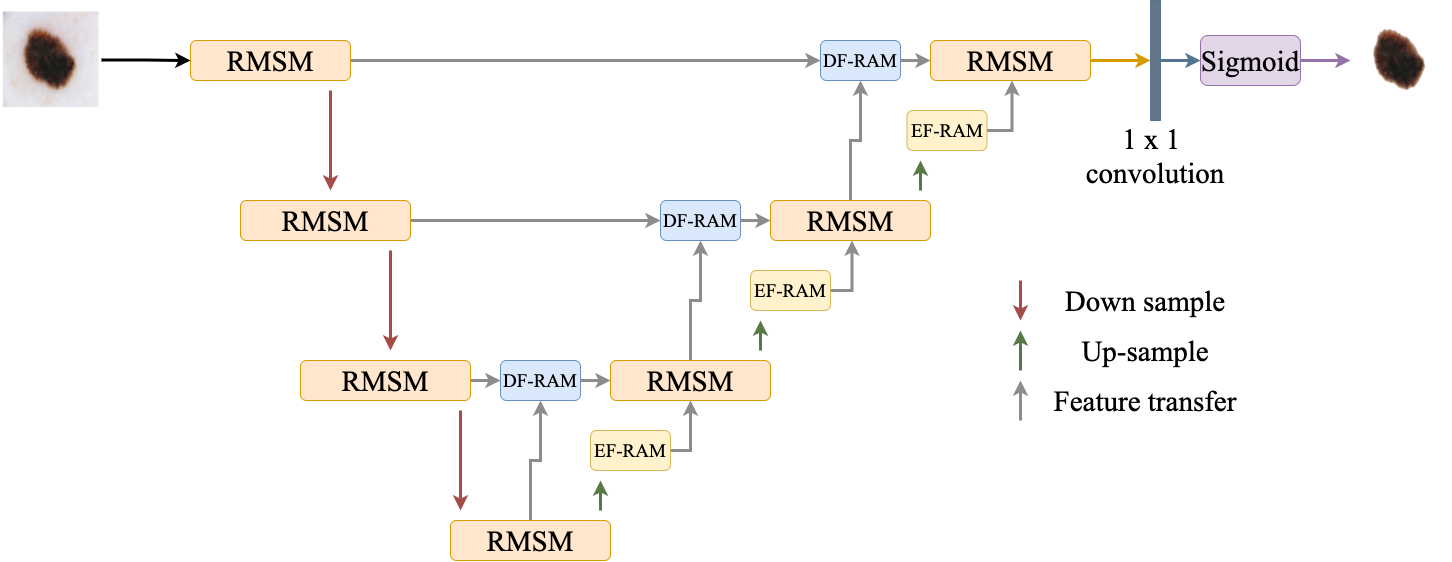}
    \caption{The proposed segmentation network with Residual Multi scale Feature Extraction module and dual attention mechanism.\label{pfg}}
\end{figure*}

\subsection{Loss function}
Since skin lesions have irregular lesion boundaries and variant shapes, a comprehensive loss function is needed to assist the model during the training process for achieving a rapid and stable convergence. In this regard, dice loss and focal loss are combined to form the loss function $Loss_{seg}$. The loss functions are shown in Equations \ref{dice}-\ref{folc}.

\begin{equation}
    Dice_{loss} = 1-2\frac{1+2P_{seg}Y_{seg}}{1+P_{seg}+Y_{seg}}
    \label{dice}
\end{equation}

\begin{equation}
\begin{aligned}
    Focal_{loss} = -Y_{seg}(1-P_{seg})^{\gamma}log(P_{seg})\\-P_{seg}^{\gamma}(1-Y_{seg})log(1-P_{seg})
    \label{folc}
    \end{aligned}
\end{equation}

In equations \ref{dice} and \ref{folc}, the $P_{seg}$ and $Y_{seg}$ represents the predicted segmentation map and the ground-truth. The $\gamma$ is the focussing parameter and is set as 2 as it has shown satisfactory results in \cite{t1, t2, t3}. The combined loss function used in this work is shown in Equation \ref{kw}.

\begin{equation}
    Loss_{seg} = Dice_{loss} + Focal_{loss}
    \label{kw}
\end{equation}

\subsection{Training}
In this work, two famous publicly available datasets, namely ISBI2017 \cite{A1} and ISIC2018 \cite{A2}, are used for training and validating the model. The skin lesion images had variant sizes in both datasets, so they are normalized into 224 x 224 resolution using the bicubic interpolation algorithm. The ISBI2017 and ISIC2018 consisted of 2000 and 2594 images. As small datasets create the problem of overfitting, data augmentation methods are employed to generalize the model. The employed data augmentation operations include Horizontal and Vertical flip, sharpening, and rotation (randomly from 0 to 90 degrees). By this procedure, four times the samples of the original datasets are generated for training the model.
In our experiments, the maximum epoch number is set to 250, and Adam \cite{A3} optimizer is chosen with a batch size of 16. The learning-rate is set to 0.0003, and this continues to drop by one-tenth after every 20 epochs.

\section{Experiments and Results}
\subsection{Datasets}
As mentioned, two publicly available datasets, namely ISBI2017 \cite{A1} and ISIC2018 \cite{A2} datasets are used in this work. The ISBI2017 dataset consists of a total of 2750 dermoscopic images divided into three sets: training set (2000 images), test set (600 images), and validation set (150 images). And the ISIC2018 dataset contains 2594 images divided into two sets: training set (2000 images) and test set (594 images). With data augmentation methods, the training images in ISBI2017 and ISIC2018 datasets are increased to 8000 images each.

\subsection{Performance metrics}

In this work, five evaluation metrics, including accuracy (ACC), specificity (SPE), recall (REC), dice coefficient (DC), and Jaccard Similarity Index (JSI), are used for validating the proposed work. These metrics are formulated in Equations \ref{ac}-\ref{js}.

\begin{equation}
    AC = \frac{TP+TN}{TP+TN+FP+FN}
    \label{ac}
\end{equation}

\begin{equation}
    SP = \frac{TN}{FP+TN}
    \label{spe}
\end{equation}

\begin{equation}
    REC = \frac{TP}{FP+TN}
    \label{rec}
\end{equation}

\begin{equation}
    DC = \frac{2|G\cap Y|}{|G|+|Y|}
    \label{dc}
\end{equation}

\begin{equation}
    JSI = \frac{|G\cap Y|}{|G \cup Y|}
    \label{js}
\end{equation}

In equations \ref{ac}-\ref{rec}, the $TP$ and $TN$ represent the number of pixels correctly predicted as lesion and non-lesion regions, $FP$ and $FN$ represent the number of pixels wrongly predicted as lesion and non-lesion regions. And in equations \ref{dc} and \ref{js}, $G$ and $Y$ are ground-truth and prediction. 

\subsection{Comparison with state of the art models}
\begin{figure*}
    \centering
    \includegraphics[width=\textwidth]{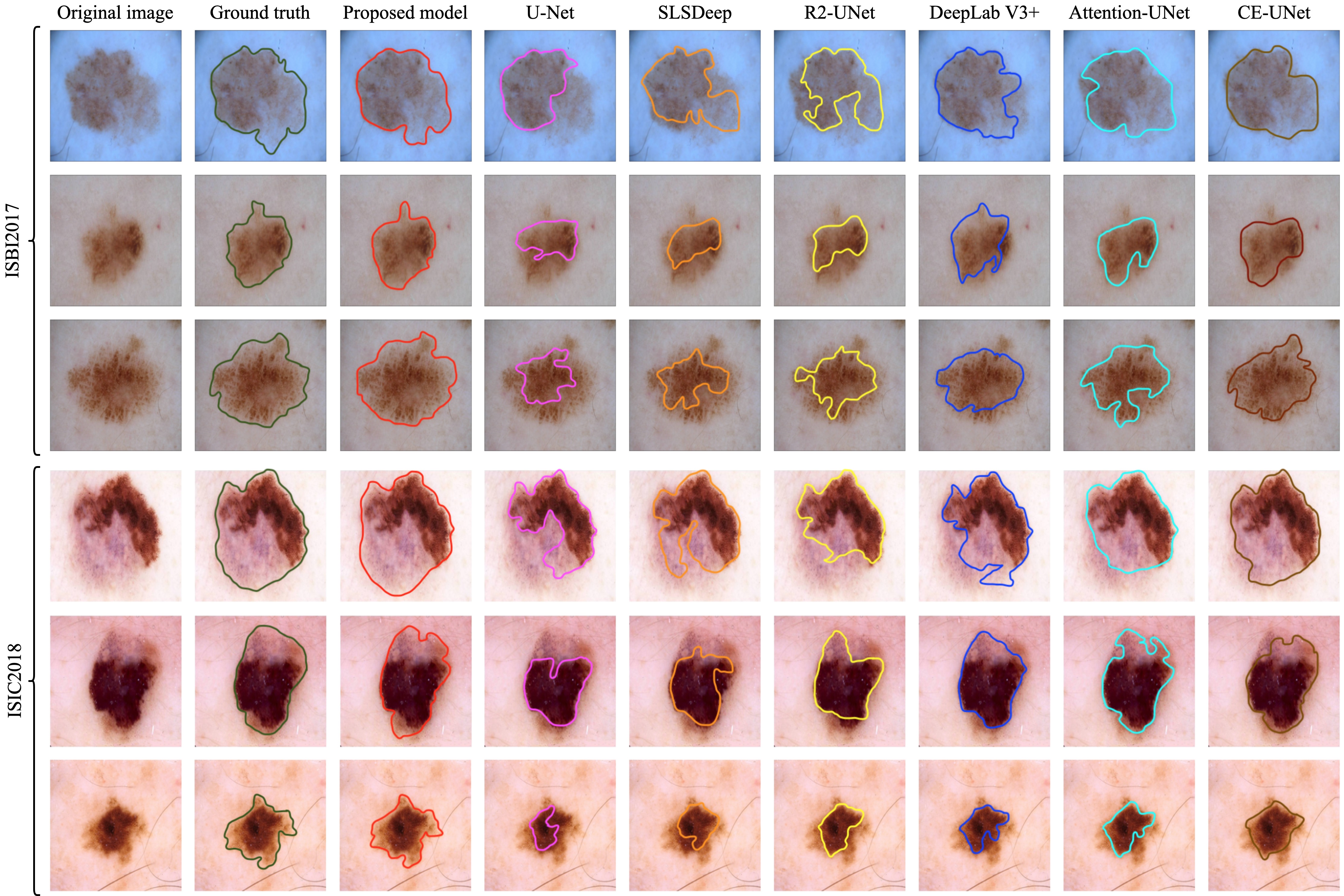}
    \caption{Comparison with the state-of-the-art methods on complex images from ISBI2017 and ISIC2018 datasets.\label{sta}}
\end{figure*}

In this work, the performance reported by the proposed model is compared with the state-of-the-art methods, including UNet \cite{W1}, CE-Net \cite{B3}, Attention-UNet \cite{I14}, DeepLab V3+ \cite{B2}, SLSDeep \cite{M14}, and R2U-Net \cite{M15}. These methods are implemented and applied to the test sets of ISBI2017 and ISIC2018 datasets using the same augmentation methods and computing environment for a fair comparison. For visual comparison, the segmentation results reported by the state-of-the-art methods and the proposed model on complex cases from both the datasets are shown in Figure \ref{sta}. It was observed that UNet is unable to accurately predict the lesion boundaries of the complex cases. With the use of pyramid pooling for enlarging the receptive fields and the combination of end-point error and log-likelihood in the loss function, the SLSDeep obtained better performance than the UNet. With the usage of recurrent, residual modules, the R2U-Net outperformed the SLSDeep model. The Attention U-Net used attention gates to aid the segmentation network in differentiating the target pixel from the background, and it performed better than earlier research. The CE-Net also reported better segmentation results using a combination of residual-multi-kernel pooling and an atrous convolution module. By employing dual attention mechanisms for enriching the decoder and up-sampled feature maps, the proposed model reported enhanced segmentation results than the existing methods. As shown in the third column of Figure \ref{sta}, the predicted boundary by the proposed model is much similar to the ground truth than the previously proposed methods, especially for complex cases with complicated color distribution and irregular boundaries.

\begin{table}[]
\caption{Comparison with the state of the art methods \label{cst}}
\centering
\begin{tabular}{@{}lllllll@{}}
\toprule
Dataset                   & Method                   & AC             & SP             & REC            & DC             & JSI            \\ \midrule
\multirow{7}{*}{ISBI2017} & U-Net                    & 88.73          & 87.84          & 79.72          & 78.45          & 71.58          \\
                          & CE-Net                   & 95.42          & 95.38          & 87.59          & 89.21          & 80.61          \\
                          & Attention-UNet           & 94.76          & 94.03          & 87.16          & 87.75          & 78.49          \\
                          & DeepLab V3+              & 93.24          & 94             & 86.64          & 87.42          & 78.05          \\
                          & R2U-Net                  & 93.12          & 93.89          & 86             & 86.73          & 77.95          \\
                          & SLSDeep                  & 92.34          & 92.17          & 85.75          & 86.30          & 76.82          \\
                          & \textbf{Proposed method} & \textbf{97.5}  & \textbf{96.94} & \textbf{94.29} & \textbf{91.16} & \textbf{83.83} \\ \midrule
\multirow{7}{*}{ISIC2018} & U-Net                    & 86.84          & 88.57          & 87.56          & 85.39          & 78             \\
                          & CE-Net                   & 95.75          & 94.32          & 94.41          & 90.26          & 84.1           \\
                          & Attention-UNet           & 94.93          & 92.59          & 93.31          & 89.45          & 83             \\
                          & DeepLab V3+              & 94.56          & 96.8           & 90.7           & 89.67          & 82.16          \\
                          & SLSDeep                  & 93.19          & 91             & 89.18          & 89.38          & 83.67          \\
                          & R2U-Net                  & 93.79          & 93.64          & 91.48          & 90             & 83.7           \\
                          & \textbf{Proposed method} & \textbf{95.92} & \textbf{97}    & \textbf{95.37} & \textbf{91.52} & \textbf{85.41} \\ \bottomrule
\end{tabular}
\end{table}

Besides the visual comparisons, the performance comparison in terms of AC, SP, REC, DC, and JSI on the ISBI2017 and ISIC2018 datasets are tabulated Table \ref{cst}. It is observed that DeepLabV3+, SLSDeep, and R2U-Net reported superior performance than the UNet in terms of accuracy. The CE-Net and Attention U-Net further enhanced the segmentation accuracy by an average of $1.85\%$ and 0.78$\%$ on ISBI2017 and ISIC2018 datasets. The proposed model reported higher performance in terms of AC, SP, REC, DC, and JSI than the existing works on both datasets, indicating the effectiveness of the proposed dual attention multi-scale model for skin lesion segmentation. As UNet is widely used for medical image segmentation, the proposed model achieved an accuracy improvement of  $8.77\%$ and $9.08\%$ on the test sets of ISBI2017 and ISIC2018 datasets.

\subsection{Ablation Study}
In this work, several ablation studies were conducted to demonstrate the utility of each module used in this work. In this regard, the performance of the proposed method is compared with baseline models including UNet, UNet with Residual Multi-scale Module (RMSM UNet), RMSM UNetwith DF-RAM, and RMSM UNet with EF-RAM. These methods are evaluated on the validation set of ISBI2017. The performance comparison is shown in Table \ref{al}.

\begin{table}[]
\centering
\caption{Results of the Ablation experiments \label{al}}
\begin{tabular}{@{}llll@{}}
\toprule
Method                                                                                    & DC             & JSI            & REC           \\ \midrule
UNet                                                                                      & 90.4           & 77.6           & 84.92         \\
RMSM Unet                                                                                 & 91.29          & 80             & 87.62         \\
RMSM UNet + DF-RAM                                                                        & 93.45          & 81.74          & 91            \\
RMSM UNet + EF-RAM                                                                        & 92.98          & 82.53          & 90.9          \\
\textbf{\begin{tabular}[c]{@{}l@{}}RMSM UNet + DF-RAM \\ +EF-RAM (Proposed)\end{tabular}} & \textbf{94.53} & \textbf{84.19} & \textbf{94.9} \\ \bottomrule
\end{tabular}
\end{table}

The UNet has several convolutional layers stacked at each level for extracting efficient features and has a skip connection, which is used to transmit information from the corresponding encoder and decoder blocks. This structural schema of the UNet acts as the backbone for the proposed work and does not have any specific mechanism for lesion segmentation, hence resulting in less segmentation JSI and DC. The concept of RMSM is to aggregate feature maps of different convolutional layers with varying kernel sizes, thus making the net wider and capable of learning more discriminative features. The residual connection is added to overcome the problem of network degradation and saturation. By adding the RMSM module, the RMSM UNet is formed, which outperformed the UNet model. The attention mechanisms DF-RAM and EF-RAM are added to refine the feature maps from the decoder and post-upsampled features. From Table \ref{al}, it is observed that RMSM UNet with EF-RAM reported higher JSI than the RMSM UNet, and RMSM UNet with DF-RAM. In contrast, the RMSM UNet with DF-RAM reported higher DC than the RMSM UNet and RMSM UNet with EF-RAM. And there is a small degradation of recall when using RMSM UNet with EF-RAM than the RMSM UNet with DF-RAM. This shows that the absence of any of the attention mechanisms will hinder the segmentation results. The inclusion of both EF-RAM and DF-RAM mechanisms in RMSM UNet yields the best performance on all the experiments. The performance increase is more than the sum of the performance boosts of each module. These experiments indicate that the introduction of EF-RAM and DF-RAM modules enhances performance mutually.

\section{Discussion}
From the presented comparative experiments and the ablation studies, it was observed that even for complex cases of skin lesions, the proposed model reported satisfactory results based on the use of multi-scale feature extraction and a dual-attention framework. From Figure \ref{sta} and Table \ref{cst}, it can be seen that the proposed model reported superior performance to the attention mechanism-based models. In addition to the comparison with the state-of-the-art works, we also compared the performance of the proposed model with the competition leaderboard presented in the Table \ref{lb}. The JSI metric is used to rank the methods \cite{B6, B5, B4} in the ISBI2017 challenge. Compared with the first ranked method \cite{B4}, the proposed model reported an enhancement of $8.23 \%$ in the JSI metric. 
For the ISIC2018 competition, the methods were also ranked based on the JSI metric. It is observed that the proposed work reported superior performance in terms of JSI and enhanced the segmentation JSI by $5.21\%$ than the first ranked method (MaskRcnn2+segmentation). On the ISIC2018 leaderboard, it was observed that several methods employed an ensemble-based approach for improvements in segmentation accuracy. These ensemble-based approaches consume more time for training, which makes them difficult to be deployed in a clinical setting. Compared with these approaches, the proposed model requires only 17 seconds to segment each image, which is very less than the existing methods.

\begin{table}[]
\begin{threeparttable}
\centering
\caption{Comparison with the top three methods from the competetion leader board \label{lb}}
\begin{tabular}{@{}lllll@{}}
\toprule
Competition               & Method                                                                                  & JSI            & AC             & DC             \\ \midrule
\multirow{4}{*}{ISBI2017} & \cite{B4}                                                                                & 75.6           & 93.4           & 84.9           \\
                          & \cite{B5}                                                                               & 76.2           & 93.2           & 84.7           \\
                          & \cite{B6}                                                                                & 76             & 93.4           & 84.4           \\
                          & \textbf{Proposed model}                                                                 & \textbf{83.83} & \textbf{97.5}  & \textbf{91.16} \\ \midrule
\multirow{4}{*}{ISIC2018} & MaskRcnn2+segmentation *                                                                & 80.2           & 94.2           & 89.8           \\
                          & Ensemble\_with\_CRF\_v3 *                                                               & 79.9           & 94.5           & 90.4           \\
                          & \begin{tabular}[c]{@{}l@{}}Automatic Skin Lesion \\ Segmentation by DCNN *\end{tabular} & 79.9           & 94.3           & 90             \\
                          & \textbf{Proposed model}                                                                 & \textbf{85.41} & \textbf{95.92} & \textbf{91.52} \\ \bottomrule
\end{tabular}
\begin{tablenotes}
      \small
      \item Note: The results in this table are obtained directly from the competition leader board. And the methods mentioned with * are not citable.
    \end{tablenotes}
  \end{threeparttable}
\end{table}

\begin{figure}
    \centering
    \includegraphics[width=\linewidth]{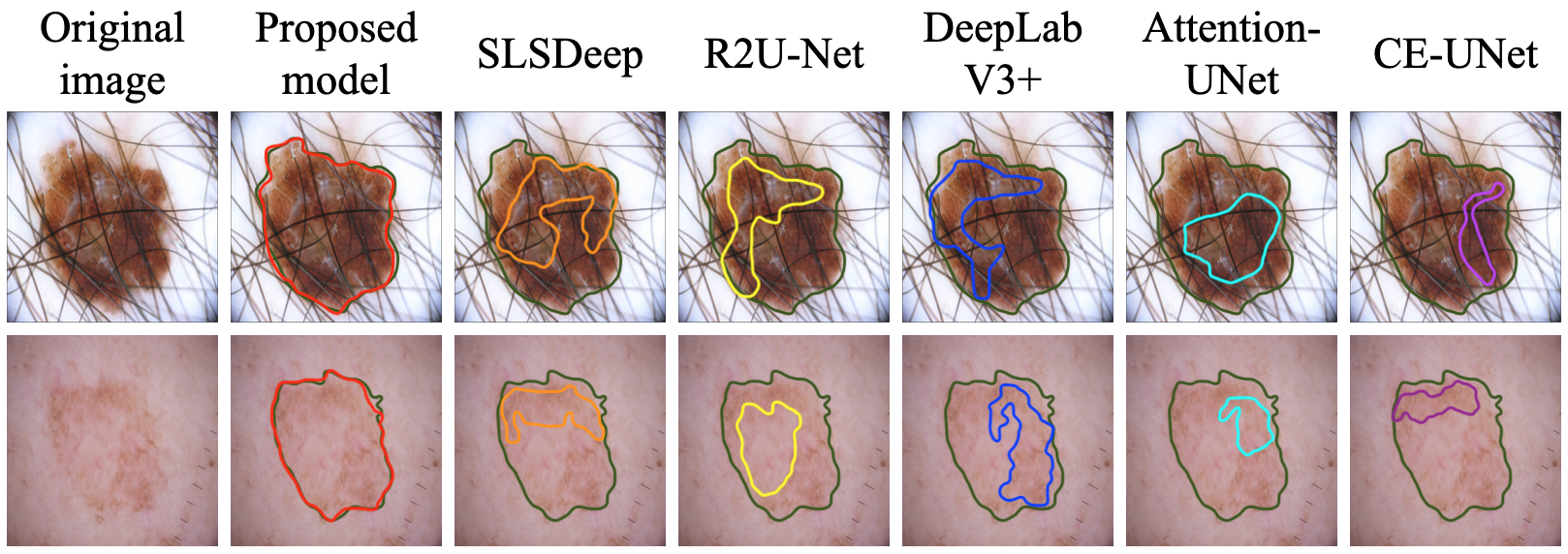}
    \caption{Comparison with the state-of-the-models on challenging cases. Orange, yellow, cyan, pink, and red contours indicate the predicted segmentation results by the state of the art models and the proposed model, and green indicates the ground truth.   \label{csop}}
\end{figure}

The ability of the proposed work in segmenting the most challenging cases is shown in Figure \ref{csop}. These cases include dermoscopic images with extremely irregular lesion boundaries, and hair in the lesion region, and the state-of-the-art methods failed on these cases, whereas the proposed model successfully segmented the lesion. The multi-scale feature extraction with the dual-attention feature refinement enabled the proposed model to deal with these challenges. The proposed model has the potential to be used in real-world scenarios for the segmentation of skin lesions from dermoscopic images. The employed dual-attention mechanism can be used in similar applications where the target region has variant characteristics of color and shape, like skin lesions.

\section{Conclusion}
In this work, a deep learning approach based on UNet is presented to segment skin lesions from dermoscopic images. Unlike the standard UNet, a new multi-scale feature extraction module is employed for extracting discriminative features to deal the challenge of skin lesion segmentation, which replaces the convolutional layers in the UNet. For enhancing the segmentation performance, a dual-attention mechanism is employed for refining the post-upsampled features and the features extracted by the encoder. This attention mechanism employs both channel and spatial attention for feature refinement. Different from the existing work, this model does not employ any post-processing or pre-processing procedures. This model is evaluated using two publicly available ISIC2018 and ISBI2017 datasets. Experimental results indicate that the proposed model outperformed the existing methods dicussed in the literature.

\bibliographystyle{amsplain}
\bibliography{bibliographty.bib}

\end{document}